\documentclass[a4paper]{article}
\usepackage{epsfig}

\usepackage{amsfonts}
\usepackage{bbm}

\begin{document}

%\doublespacing
%\flushbottom
%\raggedbottom
%\pagenumbering{roman}

\title{Natural Philosophy and Quantum Theory}\author{Thomas Marlow \\ \emph{School of Mathematical Sciences, University of Nottingham,}\\
\emph{UK, NG7 2RD}}

\maketitle

\begin{abstract}
We attempt to show how relationalism might help in understanding Bell's theorem.  We also present an analogy with Darwinian evolution in order to pedagogically hint at how one might go about using a theory in which one does not even desire to explain correlations by invoking common causes.
\end{abstract}


\section*{Motivation}
\label{sec:convergency}

In the first section we shall introduce Bell's theorem and explain how a relational philosophy might be used in interpreting it.  In the second section we will present an analogy with Darwinian evolution via two parables.  Our main aim is pedagogy---in explaining Bell's theorem to the layman.

\section*{Bell's Theorem}
\label{sec:Bell}

In his seminal work \cite{BellBOOK} Bell defines formally what he means by `locality' and `completeness' and he convincingly proves that in order to maintain his definition of `locality' one must reject his definition of `completeness' (or vice versa).  His definitions are very intuitive so they are very difficult to reject---and this is exactly the reason his theorem is so powerful.  We will outline Bell's beautiful and, we would argue, irrefutable proof below and then we shall show that using relationalism we can, as Bell proves we should, reject his definition of `completeness' and/or `locality'.  Bell's theorem is even more difficult to dissolve than Einstein, Podolsky and Rosen's famous theorem \cite{EPR} because we have, strictly speaking, to reject both his particular `completeness' assumption and his particular `locality' one---this is merely because his `completeness' assumption is an integral part of his `locality' assumption.  But we are getting ahead of ourselves.  Let us briefly discuss Bell's theorem and then we will discuss relationalism (note that we emphatically do not refute his proof because it is, in our opinion, wholly and profoundly correct---the aim of this note is to explain why this is the case).

If we have two experiments, labeled by parameters $a$ and $b$, which are spacelike separated, where $A$ and $B$ parameterise the particular results that we might receive in $a$ and $b$ respectively, then Bell assumes the following:

\begin{equation}
p(AB \vert ab I) = p(A \vert a I)p(B \vert b I)
\label{locality}
\end{equation}

\noindent  where $I$ is any other prior information that goes into the assignment of probabilities (this might include `hidden variables' if one wishes to introduce such odd things).  $A,B,a$ and $b$ can take a variety of different values.  Bell called this factorisation assumption `local causality' with the explicit hint that any `complete' theory which disobeys this assumption must embody nonlocal causality of a kind.

Jarrett \cite{Jarrett84} showed that this factorisation assumption can be split up into two logically distinct assumptions which, when taken together, imply Eq.\,(\ref{locality}).  These two assumptions are, as named by Shimony \cite{Shimon86}, parameter independence and outcome independence.  Shimony \cite{Shimon86} argued that orthodox quantum theory itself obeys parameter independence, namely:

\begin{equation}
p(A \vert abI) = p(A \vert aI).
\label{parameter}
\end{equation}

\noindent
This means that, in orthodox quantum theory, the probability that one predicts for an event $A$ in one spacetime region does not change when the parameter $b$ chosen in measuring an entangled subsystem elsewhere is known. Knowledge of $b$ doesn't help in predicting anything about $A$.  If parameter independence is disobeyed then, some suggest \cite{Shimon86}, we would be able to signal between spacelike separated regions using orthodox quantum theory and, as such, there seems to be no harm in presuming that parameter independence failing means that nonlocal causality is probably manifest.  However, by far the more subtle assumption implicit in Bell's `locality' assumption is outcome independence (which orthodox quantum theory does not obey):

\begin{equation}
p(A \vert BabI) = p(A \vert abI).
\label{outcome}
\end{equation}

\noindent
Knowledge of $B$ can and does effect the predictions one makes about $A$ in the standard interpretation of quantum theory \cite{Shimon86}.  Bell implicitly justified assumption (\ref{outcome}) by presuming `completeness' of a certain kind.  If one has `complete' information about the experiments then finding out the outcome $B$ cannot teach you anything about the probability of outcome $A$, hence we should presume (\ref{outcome}) for such `complete' theories.  So outcome independence does not necessarily have anything to do with locality.  Even if one were a nonlocal observer who gathered his information nonlocally one would still assume (\ref{outcome}) as long as that information was `complete'.  So two assumptions go into Bell's factorisation assumption, namely a `locality' assumption (\ref{parameter}) and a `completeness' assumption (\ref{outcome}).  Bell then proves elegantly that (\ref{parameter}) and (\ref{outcome}) together imply an inequality that quantum theory disobeys.  So, quantum theory does not obey Bell's locality condition (\ref{locality}) which is itself made up of two sub-assumptions (\ref{parameter}) and (\ref{outcome}).  Note that \emph{the one sub-assumption that orthodox quantum theory emphatically does not obey is the `completeness' one} (\ref{outcome}).  We hope to convince you that any justification of (\ref{outcome}) relies on a category error\footnote{One argument against this presentation \cite{MaudlinBOOK} is that the splitting of the Bell's factorisation assumption into (\ref{parameter}) and (\ref{outcome}) is not unique, and one could equally well split the factorisation assumption into different sub-assumptions.  There is no \emph{a priori} reason to choose one particular way to split (\ref{locality}) into prior assumptions.  This we agree with, but note that there are \emph{a posteriori} reasons from choosing to use Jarrett's analysis while discussing orthodox quantum theory.  Since we are not refuting Bell's theorem---we would rather merely adopt a different interpretation of his assumptions---it does not matter which way we split (\ref{locality}), we merely split it up into (\ref{parameter}) and (\ref{outcome}) for pedagogical reasons.}.

So we used completeness to justify demanding (\ref{outcome}) but could we not similarly have used completeness to demand (\ref{parameter})?  Yes we could.  So the problem arises that it is difficult even to justify (\ref{parameter}) as a locality assumption.  The only reason we \emph{do} call (\ref{parameter}) a locality assumption is that some theories which disobey it could perhaps be used to signal between spacelike separated regions.  The point, however, is mute because quantum theory (in the orthodox interpretation) obeys (\ref{parameter}) and hence we have no problem with signalling in standard quantum theory \cite{Shimon86}.  So Bell's theorem seems to suggest that quantum theory is `incomplete' rather than `nonlocal'.  The very assumption that we should demand that we can ever find the \emph{real, absolute, ontological} probability independently of the probabilities we assign to other propositions seems to be \emph{the} assumption that orthodox quantum theory disobeys in Bell's analysis.  Hence one might wish to adopt a relational approach to probability theory \cite{Marlow06fold} where we cannot even define the probability of an event independently of the probabilities we assign to other events---\emph{cf.} Cox's axioms of probability \cite{CoxBOOK,Cox46,JaynesBOOK}.  So we must, it seems, try and interpret quantum theory in such a way that demanding (\ref{outcome}) is simply a category error of some sort (rather than something we desire in the first place)---Bell's theorem is so elegant and irrefutable that we must try to accommodate his conclusions.

Relational theories are ones in which we do not use statements about mere ontology, we only use facts or definitions about things that we can rationally justify.  In fact, this is the only real principle of relational theories introduced by Leibniz, it is called the Principle of Sufficient Reason (PSR) \cite{Smolin05}.  The PSR implies another principle called the Principle of Identifying the Indiscernible (PII) which says that if we cannot distinguish two theoretical entities by distinctions that we can rationally justify, we should identify them.  In other words, if two things are to be distinct then they must be distinct for an identifiable reason.  No two equals are the same.  These are desiderata that ensure that we do not introduce facts, statements or distinctions in our theories which we are not rationally compelled to.  However natural or obvious our assumptions sound, if we are not rationally compelled to use them we clearly ought not to use them in case we are mistaken (and, of course, there are lots and lots of ways we might be mistaken).  So, for example, it seems very natural and obvious to define motion with respect to an absolute `flat' space and time (like in Newton's theory of gravity) because we instinctively na\"{\i}vely feel this is the case but it turns out that, since we are not rationally compelled to use such an absolute definition, we ought not to.  This was Einstein's insight and it allowed him to derive a better theory than Newton's.  In \cite{Marlow06fold} we argued that such a `parsimonious' relational philosophy can help in interpreting and generalising probability theory and quantum theory.  For other relational approaches to quantum theory see \cite{Rovel96,SR06,Marlow06weak}.

\section*{Convergency}

It may seem like an odd jump to now begin discussing an analogy with Darwinian evolution.  Nonetheless, this is exactly what we shall do and let us explain why.  It is easy to see that Darwin also seemingly used such careful rational desiderata \cite{Smolin05}.  Many people used to claim that species were eternal categories.  One might perversely say that species possessed `elements of biology' that could not change.  Rabbits are rabbits because they embody some eternal `Rabbitness'.  A few hundred years ago this assumption didn't seem too implausible.  Nonetheless, it is wrong.  Utterly fallacious.  Darwin argued that one was not rationally compelled to make such an assumption.  And rejecting this assumption allowed him to design, with considerable effort and insight, his theories of natural and sexual selection.  It might be that an analogous rationalist approach might help in understanding Bell's theorem.

Inspired by \cite{DawkinsBOOK}, let us introduce a little parable.  Imagine that, in some horrible future, we find two species which look and behave in identical ways and these two organisms each call home an island that is geologically separate from the other.  This amazes us so we search as hard as we possibly can but we find that, beyond reasonable doubt, that there is \emph{no} way that a common ancestor of the two species could have recently (in geological time) `rafted', `tunneled', flown or swam between the two islands.  This amazes us even more. Furthermore, we check the two organisms' genetics and we find that they each have \emph{exactly the same genome}.  Exactly the same.  We must conclude that two species have evolved separately and have exactly the same genome.  This more than merely amazes us, it shocks us to our cores, and we reel about jabbering for years trying to understand it.  We know of no (and cannot rationally justify a) physical mechanism that can ensure a common genome for separate species.  Some biologists commit suicide, some find God, some continue to search for a way that the two species might have evolved and conspired together regardless of the fact they are separate, and others consider it a fluke and move on. However, I am confident that we do not need to worry about this parable, this future will never become reality... beyond reasonable doubt anyway.

Note that this parable suggests a reading that is \emph{exactly opposite} to many expositions of Bell's theorem (especially Mermin's `instruction set' approach, see \cite{Mermin02} and references therein).  One might question whether the `instruction sets' (or, equivalently, the catalogue of causes) we assign to two separated correlated particles are common to both.  If we assume the instruction sets for each particle should be the same for two correlated particles then we conclude that quantum theory is surprising (it disobeys this assumption or embodies causal nonlocality).  Our parable suggests the opposite view.  Turning the argument on its head we should be \emph{utterly shocked} to find that the `instruction sets' or catalogues of causes are the same in each region.  We know of no (and cannot rationally justify a) physical mechanism that can ensure common causes within the separate regions.  We are shocked if two correlated phenotypes evolved separately and ended up embodying the same genetics, but why would we not be shocked if two correlated particles were shown to have arisen from common causes.  The orthodoxy \emph{demands} that separate correlated properties should arise from common causes.  Why the distinction between the two analogous cases?  Why?

But, of course, we need another parable because perhaps the two organisms embody the same phenotypes merely because they live in sufficiently similar surroundings, so it might be that the `instruction sets' or common causes arise in the environment of the two separate convergent species.  So we use two different definitions of `niche': an organism has its internal `genetic' niche and it has its external `environmental' niche.  Convergent species might evolve in geological separation seemingly \emph{because} their environmental niches are sufficiently similar.  But just as two distinct genetics can give rise to the same phenotype, might it not be that two distinct `ecological' niches might also house a common correlated phenotype?  Yes!---this happens all over the place; biology is abundantly and wonderfully filled with convergent phenotypes that exist within significantly distinct environmental niches. (Species like Phytophthora, Platypus and Thylacine---Tasmanian Tiger---are all good examples of species which seem to embody convergent phenotypes with respect to other separated species; similarly, Koalas have fingerprints...).  So we begin to question whether, in biology, we should ever demand that correlated phenotypes arise from common causes. Certainly we might happily begin to question such a thing without resorting to proclamations about nature being `nonlocal'.

So, another parable.  Imagine in the future we again have two geologically separate islands and two seemingly identical species.  Lets say they are apes, and that all the phenotypes that we can identify are pretty much the same: they have the same gait, both have grey hair on their heads, squat when they defaecate, speak the same jargon, the same kidney function etc.  We would be very surprised if they have the same genetics as long as we rule out the idea that a recent ancestor `rafted' between the islands.  We catch these apes and try to make them mate, but they do not seem to want to.  We talk to them nicely and even though it disgusts them they mate.  They do not produce any offspring or, if they do, their offspring are sterile\footnote{This is a biological version of the relational Principle of Identifying the Indiscernible.}; they are different species with different genetics.  We are happy about this because it confirms that they evolved separately as we would expect two geologically separate species would.  But then we ask ourselves, why do they share common phenotypes regardless of the fact they evolved in geologically separate niches?  Perhaps they share common phenotypes because their niches are sufficiently similarly simian.  So now let us list common features of their ecological niches:  the forests on each islands have similarly shaped trees, a similar array of fruit, a similar array of predators, the temperatures and humidity are the same, the same sun shines down on the two islands, etc.  So it seems that the two apes have evolved common phenotypes \emph{because} their niches embody common causes.  This seems obvious, right?  Bell convincingly, and profoundly, proves that we cannot use analogous sets of common causes to explain correlations in quantum theory while maintaining causal locality, for such a `complete' local theory would obey (\ref{outcome}).

However, note that, in biology at least, we have exactly the same problem we had with the `genetics' parable.  We should look at it from \emph{exactly the opposite point of view}.  If we were to find that the environmental niches were so amazingly similar between the islands so as to ensure that two simians existed with the same phenotypes we would be utterly shocked.  If the local ecologies had `genomes'---we might call them `e-nomes'---we would be shocked if their `e-nomes' were the same (for \emph{exactly the same} reason we would be shocked if two species that have evolved separately for a long time have the same genome).  It would be as if the environments of each island were conspiring with each other, regardless of geological separation, so as to design simian apes.  \emph{That} would be, by far, the more amazingly shocking conspiracy in comparison to what, in fact, actually happens: convergent phenotypes evolve with distinct genetics and within distinct ecologies.  Common causes are the conspiracy---an analogue of the illusion of design.  So we must, in biology, be careful to distinguish common causes from `sufficiently similar' uncommon causes which might arise by natural means.  We simply \emph{cannot justify} the presumption of common causes.  Like a magician on stage, nature does not repeat its tricks in the same way twice\footnote{The space of sets of possible causes for a particular phenotype to be the case is so vast that nature is unlikely to use the same set of causes the next time.  There are many ways to \emph{evolve} a cat.}.  Bell's theorem proves that we cannot assume an analogous common cause design-conspiracy in quantum theory while maintaining causal locality, but clearly it is a logical possibility that, however unpalatable, we ought \emph{not to assume it in the first place}.  So, just like Bohr \cite{Bohr35} famously rejected EPR's \cite{EPR} assumptions and dissolved their nonlocality proof, we might yet happily reject Bell's \cite{BellBOOK} assumptions.  And Bell (like EPR) \emph{proved} that we should reject his (their) assumptions.  Two interpretations remain: some suggest that we should search for a way to maintain causal locality \cite{SR06,Bohr35,Peres03,Jaynes89,BellBOOKsub} while others suggest that we should use causally nonlocal common cause theories because they are, in the least, pedagogical and easy to understand \cite{BellBOOK}.

Furthermore, we can ask ourselves another penetrating question:  in our analogy we know that the islands are similar, that they have similar forests, fruits, predators, temperatures and so forth but what, by Darwin's beard, gives us the sheer audacity, the silly ape-centric irrationalism, to call such things `causes'?  Such a thing stinks of teleology.  The `real' causes are small unpredictable changes by seemingly `random' mutation.  Niches don't `cause' particular phenotypes to be the case, neither as a whole nor in their particulars.  Analogously, what sheer teleological audacity do we have to discuss hidden `causes' in quantum theory?  Measurements don't cause particular properties to be the case.  Phenotypes and properties merely evolve---and we can design theories of probabilistic inference in which we assign probabilities to propositions about whether such phenotypes (resp. properties) will be said to be the case in certain niches (resp. measurements)\footnote{This Darwinian analogy correlates quite well with the idea of complementarity \cite{BohrBOOK}.  `Complimentary' phenotypes rarely arise within the same niche merely because they tend to arise in different niches.}.  Nonetheless we don't have to give up on causality, nor do we have to give up local causality (merely some of our mere definitions of `local causality'). Local causality ought to be inviolate unless we find something physical, that we can \emph{rationally justify}, that travels faster than light.

This analogy with biology, and Bell's theorem itself, suggests that common causes might be something that we \emph{desire} but they are not something that we can \emph{demand} of nature;  common causes are an anthropomorphic ideal.  So, now let us ask ourselves, where might this analogy fail?   Correlations between separate phenotypes arise over long periods whereas quantum correlations arise over very short time scales.  Perhaps convergency will not have enough `time' to occur in quantum theory.  However, `long' is also defined relative to us: the statutory apes.  Quantum correlations happen over `long' periods of `Plank' time.

There is one interesting way in which the analogy succeeds brilliantly.  Evolution happens by ensuring that small changes of phenotypes occur unpredictably.  Quantum mechanics is, at its very core, a theory of small\footnote{\emph{Cf.} Hardy's `damn classical jumps' \cite{Hardy01}.} unpredictable changes as well.  Perhaps we can learn lessons from quantum theory that are similar to those we learnt from Darwin.  Rationally we cannot justify any conspiracy or design (some demon or god which ensures correlations arise from common causes) so rather, we must invoke a theory that explains all the `magical' and `interesting' correlations we see in nature by manifestly rejecting such accounts. Instead we should search for a physical mechanism by which convergency occurs in a causally local manner (interestingly, this is an approach suggested by Bell himself\footnote{He discusses trying to define a more `human' definition of locality that is weaker than his formal definition---we would argue, in opposition, that his formal definition (\ref{locality}) is the `human' one because it panders to an anthropomorphic ideal and is \emph{clearly not obeyed by nature.}} in \cite{BellBOOKsub}).  Let us learn from life.

So, this brings us back to relationalism.  If we follow a relational philosophy we need to obey the Principle of Identifying the Indiscernible (PII).  Remember that the PII ensures that if all the facts are telling us that two things are exactly the same then we must identify them, they are the same entity and not separate entities.  Let us catalogue all the `causes' for one event to be the case, or even for it to probably be the case (one might call these `causes' hidden variables).  Similarly for another correlated event at spacelike separation.  The PII tells us that these two catalogues of `causes' \emph{must not be the same}.  The catalogues of `causes' cannot be the same otherwise we would identify indiscernibles and we could not be discussing separate entities. This \emph{uncommon} `cause' principle should not worry us---it is the very definition we use to allow us to call two things separate in the first place.  This absolves us of reasoning teleologically about nonlocality.  Instead we can reason rationally about locality.  So two species that evolve separately will, in general, be separate species that cannot breed however similar the phenotypes that they embody.  If we isolate two species, we know of no natural mechanism which maintains common causes in geological time, there are no `elements of biology'.  Analogously, two separate correlated particles are, in fact, separate \emph{because} they arise from uncommon `causes'.

\section*{Summary and Conclusion}

Bell's theorem takes the form: `something' plus casual locality logically implies a condition---let's call it $C$---and quantum theory disobeys $C$.  Rather than fruitlessly argue over whether Bell's theorem is logically sound or whether quantum theory disobeys $C$, we have tried to discuss why people choose to reject causal locality rather than the `something else' that goes into Bell's theorem.  Bell was quite explicit in noting that the reason he chose to reject causal locality was because the `something' constituted some very basic ideas about realism that he didn't know how to reject.  That `something' might be called `completeness' and we obviously don't want quantum theory to be incomplete.  Hence Bell's conclusions rely on that `something' being eminently desirable---a very \emph{anthropomorphic} criteria.  Also, he noted that we are used to causally nonlocal theories (\emph{cf.} Newtonian gravity) and that they are pedagogically useful and easy to understand.  These are all good, if not wholly compelling, reasons for choosing to reject causal locality.

The major problem for those who would rather not reject causal locality (until a particle or information is found to travel faster than light) is in identifying clearly what that `something' is, or what part of that `something' we should reject.  We have used a relational standpoint to give two options.  Clearly Bell made some very specific assumptions about probability theory and it might be here that we can nullify his theorem.  Perhaps probability theory ought to be relational, and thus it is not clear whether we should demand outcome independence (\ref{outcome}) on logical grounds \cite{Jaynes89}.  If we define probabilities relationally then what we \emph{mean} by a probability will be its unique catalogue of relationships with all other probabilities (regardless of any separation of the events to which those probabilities refer).  This doesn't seem to convince people, perhaps because Bell's justification for (\ref{locality}) didn't seem to come from probability theory \emph{per se} but rather it stemmed from some notion of `completeness' or `realism' (an argument which falters when we note that Bell's ideas of `realism' or `completeness' might themselves stem from a particular understanding of probability theory \cite{Jaynes89}). Nonetheless we have provided a second way out.  Instead of assuming that the pertinent part of that `something' is mainly to do with probability theory, let us go to the heart of Bell's theorem and assume that `something' that we might yet reject is ``correlations ought to have common causes''.  This common cause principle is the cornerstone of Bell's theorem, but perhaps it is just plain wrong.  Like `elements of biology' are just plain wrong.

We have given an example of a physical theory---if also a biological theory---which convincingly inspires doubt about the anthropomorphic desire for common causes.  This is not to suggest that Darwinian evolution necessarily disobeys $C$, nor that we necessarily ought to use Darwinian principles in quantum theory.  Rather, ``correlations ought to have common causes'' might be a rejectable assumption, and there might be \emph{good reason} to reject it.  If one agrees with a relational philosophy then there is a simple argument that suggests that correlations ought to have \emph{un}common causes, \emph{cf.} Leibniz's PII.  Even if you do not accept this na\"{\i}ve argument, we hope that the analogy with Darwinian evolution will, in the least, convince you that the assumption of common causes might yet \emph{possibly} be rejected by rational reasoning (there are theories out there, even if quantum theory is not yet one of them, where the presumption of common causes for correlations just doesn't hold true, and for good reason).  Perhaps we have even convinced you that uncommon causes are \emph{plausible}, and that there might yet be found some natural mechanism by which quantum-convergency occurs---one that we might yet identify and investigate.  It is often suggested that causal nonlocality is a route that we logically need not take, but it is also possible that we \emph{ought} not to take it.

\section*{Acknowledgements}

We owe no small debt to the late E. T. Jaynes' work---he seemed to understand Bell's theorem even without the use of a convenient analogy.  Thanks are also due to EPSRC for funding this work, Lucien Hardy for a critical reading of a first draft and Matthew Leifer for comments on a second draft.


\begin{thebibliography}{99}

\bibitem{BellBOOK} J. S. Bell, \emph{Speakable and unspeakable in quantum mechanics}, (Cambridge University Press, 2004).

\bibitem{EPR} A. Einstein, B. Podolsky and N. Rosen, ``Can Quantum-Mechanical Description of Physical Reality Be Considered Complete?''  \emph{Phys. Rev.} \textbf{47} (1935) 777.

\bibitem{Jarrett84} J. P. Jarrett, ``On the Physical Significance of the Locality Conditions in the Bell Arguments'' \emph{No\^{u}s} \textbf{18} (1984) 569--589.

\bibitem{Shimon86} A. Shimony, ``Events and Processes in the Quantum World'', in \emph{Quantum Concepts in Space and Time}, Penrose, R. and Isham, C. J. eds. Clanderon Press, Oxford, 1986.

\bibitem{Marlow06fold}  T. Marlow, ``Into the Fold: Searching for a Theory of Natural Inference'' (2006) preprint: {\tt quant-ph/0605142}.

\bibitem{MaudlinBOOK} T. Maudlin, \emph{Quantum Nonlocality and Relativity} (Blackwell, 1994).

\bibitem{CoxBOOK} R. T. Cox, \emph{The Algebra of Probable Inference} (The Johns Hopkins University Press, 1961).

\bibitem{Cox46} R. T. Cox, ``Probability, frequency, and reasonable expectation'' \emph{American Jour. Phys.} \textbf{14} (1946) 1--13.

\bibitem{JaynesBOOK} E. T. Jaynes, \emph{Probability Theory: The Logic of Science} (Cambridge University Press, 2003).

\bibitem{Smolin05}  L. Smolin, ``The case for background independence'' (2005) preprint: {\tt hep-th/0507235}.

\bibitem{Rovel96} C. Rovelli, ``Relational Quantum Mechanics'' \emph{Int. Jour. Theo. Phys.} \textbf{35} (1996) 1637--1678, preprint:  {\tt quant-ph/9609002}.

\bibitem{SR06} M. Smerlak and C. Rovelli, ``Relational EPR'' (2006) preprint: {\tt quant-ph/0604064 v1}.

\bibitem{Marlow06weak}  T. Marlow, ``Weak Values and Relational Generalisations'' (2006) preprint: {\tt gr-qc/0604085}.

\bibitem{DawkinsBOOK} R. Dawkins, \emph{The Ancestor's Tale} (Phoenix Press, 2005).

\bibitem{Mermin02} N. D. Mermin, ``Shedding (red and green) light on``time related hidden parameter'''' (2002) preprint: {\tt quant-ph/0206118}.

\bibitem{Bohr35} N. Bohr, ``Can Quantum Mechanical Description of Physical Reality be Considered Complete?'' \emph{Phys. Rev.} \textbf{48} (1935) 696--702.

\bibitem{Peres03} A. Peres, ``Einstein, Podolsky, Rosen, and Shannon'' \emph{Found. Phys.} \textbf{35} (2003) 511--514, preprint: {\tt quant-ph/0310010}.

\bibitem{Jaynes89} E. T. Jaynes, ``Clearing Up Mysteries\----The Original Goal'' in \emph{Maximum Entropy and Bayesian Methods}, J. Skilling ed., Kluwer Academic, Dordrecht, 1989.

\bibitem{BellBOOKsub} J. S. Bell, ``The theory of local beables'' in \cite{BellBOOK} p. 52--62.

\bibitem{BohrBOOK} N. Bohr, \emph{Atomic Physics and Human Knowledge} (New York, John Wiley and Sons Inc., 1958).

\bibitem{Hardy01} L. Hardy, ``Quantum Theory From Five Reasonable Axioms'' (2001) preprint: {\tt quant-ph/0101012 v4}.

All preprints refer to the http://arxiv.org/ website.

\end{thebibliography}
\end{document}